\documentclass[12pt, A4]{article}
\usepackage{amssymb}
\usepackage{amsmath}
\usepackage{amsfonts}
\usepackage[dvips]{graphicx}
\usepackage{setspace}
\usepackage{latexsym}
\usepackage{epsf}
\usepackage{epsfig}
\usepackage{braket}
\usepackage{slashed}
\usepackage{color}
\usepackage{hyperref}
\hypersetup{
    colorlinks=true, %set true if you want colored links
    linktoc=all,     %set to all if you want both sections and subsections linked
    linkcolor=black,  %choose some color if you want links to stand out
    pdfstartview={XYZ null null 1.00},
}
\begin{document}
\vspace*{\fill}
\begin{center}
\textbf{\Large A Light Dilatonic Higgs and a Dilatonic Inflaton in the Georgi-Glashow SU(5) Model}
\end{center} 
\vspace{0.2 cm}
\begin{center}
\textbf{Hassan Saadi} \\
\vspace{0.1 cm}
\textit{Mathematics Department and Institute of Applied Mathematics, University of British Columbia, Canada}
\end{center}
\vspace{1.5 cm}
\begin{abstract}
\vspace{0.3 cm}
In this paper, two dilatonic fields that capture two conformal symmetries in the Georgi-Glashow $SU(5)$ model are introduced. Then, one of the fields is associated with the Higgs, and the other one with inflation. By using the Higgs portal, the mass of the dilatonic Higgs can be tuned by the coupling constant between the two dilatonic fields. 
\end{abstract}

\section{Introduction}
An explanation for the existence of a light Higgs is still missing. Various approaches have been proposed. One candidate is a dilaton: a Goldstone Boson (GB) that emerges when conformal invariance is spontaneously broken. In a strongly coupled electroweak symmetry breaking (EWSB) sector, the conformal breaking scale $f$ is not necessarily equal to $\nu\simeq 246 GeV$. When the dilaton is realized non-linearly, it couples to the Standard Model (SM) at the classical level similar to the Higgs \cite{Goldberger}.  For more literature review about dilatons and scale invariance in particle physics \cite{Literature}. \\
In this paper, a dilatonic field $\phi$ is added as a conformal compensator to make the $SU(5)$ Lagrangian conformally invariant. We will have a potential $V(\phi)$ for $\phi$, in addition to the potential $V(\chi)$ that is associated with the dilaton field $\chi$ that makes the electroweak chiral Lagrangian conformally invariant as in \cite{Goldberger}. A potential term $V(\phi,\chi)$ is introduced, and it represents couplings between the two dilatonic fields and respects discrete symmetry $\phi\leftrightarrow -\phi$, $\chi \leftrightarrow -\chi$. Once $\phi$ picks up a vacuum expectation value, the dilatonic Higgs would acquire a mass through the Higgs portal \cite{Higgsportal1}\cite{Higgsportal2} for a given coupling constant. \\
In the first section, we review how the dilatonic field $\chi$ couples to the electroweak chiral
Lagrangian, then demonstrate how the dilatonic field $\phi$ couples to the $SU(5)$ Lagrangian. In the second section, a potential is suggested that shows how the dilatonic Higgs can acquire a small mass through the Higgs portal.
\section{Couplings to the Electroweak Chiral Lagrangian and SU(5)}
A field $\chi(x)$ is introduced as a conformal compensator to make a Lagrangian conformally invariant. Consider a general Lagrangian in the basis of anomolous dimension eigenoperators $\mathcal{L}=\sum_{i} g_{i}(\mu)\mathcal{O}_{i}(x)$. If $\lambda: x^{\mu} \rightarrow e^{\lambda}x^{\mu}$, then the eigenoperators should transform as $\mathcal{O}_{i}(x) \rightarrow e^{\lambda d_{i}}\mathcal{O}_{i}(e^{\lambda}x)$ where $[\mathcal{O}_{i}]=d_{i}$. Hence, $\lambda: \chi(x) \rightarrow e^{\lambda} \chi(e^{\lambda}x)$, and we can generalize it in $\mathcal{L}$ as $g_{i}(\mu) \rightarrow g_{i}\Big(\mu\frac{\chi}{f} \Big)\Big(\frac{\chi}{f}\Big)^{4-d_{i}}$. Let $f=\braket{\chi}$ be the order parameter where conformal symmetry breaks. Conformal symmetry breaking has a GB $\chi(x)=fe^{\sigma(x)/f}$ where $\sigma(x)$ transforms non-linearly as $\frac{\sigma(x)}{f} \rightarrow \frac{\sigma(e^{\lambda}x)}{f}+ \lambda$. \\
In order to make the electroweak chiral
Lagrangian conformally invariant at the classical level, a conformal compensator $\chi/f$ with the appropriate power multiplies the terms of the Lagrangian as \cite{Goldberger}\cite{Coleman}
\begin{equation}
\mathcal{L}_{\chi SM}=\Big(2\frac{\bar{\chi}}{f}+\frac{\bar{\chi}^{2}}{f^{2}}\Big)\Big[m_{W}^{2}W_{\mu}^{+}W^{-\mu}+\frac{1}{2}m_{Z}^{2}Z_{\mu}Z^{\mu}\Big]+ \frac{\bar{\chi}}{f}\sum_{\psi}m_{\psi}\bar{\psi}\psi \label{LchiSM}
\end{equation}
where $\bar{\chi}(x)=\chi(x)-f$ and we expanded about $\braket{\chi}=f$. Hence, to make the Lagrangian conformally invariant, we make the substitution $\nu\rightarrow\nu\frac{\chi}{f}$. In $SU(5)$, we incorporate a conformal compensator $\phi/F$, where $F$ is the conformal breaking scale, as
\begin{equation}
\mathcal{L}_{\phi, SU(5)}= \Big(2\frac{\bar{\phi}}{F}+\frac{\bar{\phi}^{2}}{F^{2}}\Big)\Big[m_{X}^{2}X_{\mu}X^{\mu}+m_{Y}^{2}Y_{\mu}Y^{\mu}\Big] \label{LphiSU5}
\end{equation}
where $X$ and $Y$ are leptoquarks with masses $m_{X}=m_{Y}=\frac{5g_{5}}{2\sqrt{2}}\nu_{24} \sim 10^{15} GeV$, and $\bar{\phi}(x)=\phi(x)-F$ and we expanded about $\braket{\phi}=F$. In order to make the Lagrangian (\ref{LphiSU5}) conformally invariant, we can make the substitution $\nu_{24} \rightarrow \nu_{24}\frac{\phi}{F}$. Note that both $\chi$ and $\phi$ have self-interactions terms of the form
\begin{eqnarray}
\mathcal{L}_{\chi} &=& \frac{1}{2}\partial_{\mu}\chi\partial^{\mu}\chi+\frac{c_{4}}{(4\pi\chi)^{4}}(\partial_{\mu}\chi\partial^{\mu}\chi)^{2}+... \\
\mathcal{L}_{\phi} &=& \frac{1}{2}\partial_{\mu}\phi\partial^{\mu}\phi+\frac{d_{4}}{(4\pi\phi)^{4}}(\partial_{\mu}\phi\partial^{\mu}\phi)^{2}+... 
\end{eqnarray}
where the constants $c_{4},d_{4} \sim \mathcal{O}(1)$ are determined by the underlying CFT by the a-theorem \cite{atheorem1} \cite{atheorem2}.
\section{A Light Dilatonic Higgs in SU(5)}
Let us propose a potential of the form
\begin{equation}
\tilde{V}(\phi,\chi)= V(\phi)+V(\phi,\chi)+V(\chi) \label{Vtilde}
\end{equation}
where
\begin{eqnarray}
V(\phi) &=& A\phi^{4}\Bigg[\ln\Big(\frac{\phi}{F}\Big) -\frac{1}{4}\Bigg] \label{Vphi} \\
V(\phi,\chi) &=& -c\chi^{2}\phi^{2} \label{Vphichi} \\
V(\chi) &=& \frac{\lambda\chi^{4}}{4!} \label{Vchi}
\end{eqnarray}
where $A=\frac{m_{\phi}^{2}}{4F^{2}}\sim g_{5}^{4}$. The potential $V(\phi)$ is defined such as
\begin{eqnarray}
V^{\prime}(\phi)= 0 \quad \Rightarrow \quad \braket{\phi}=F \label{phiexp}\\
\frac{d^{2}V(\phi)}{d\phi^{2}}\Bigg|_{{\phi}=\braket{\phi}}= m_{\phi}^{2}
\end{eqnarray}
$V(\phi)$ in (\ref{Vphi}) is a potential for the dilatonic field $\phi$ that acts as a conformal compensator in a CFT of the leptoquarks in $SU(5)$. $V(\chi)$ in (\ref{Vchi}) is a potential for the dilatonic field $\chi$ that acts as a conformal compensator in a CFT of the SM. And finally, $V(\phi,\chi)$ in (\ref{Vphichi}) is a term that couples $\phi$ and $\chi$ which is called the Higgs portal with a negative sign.\\
Substituting (\ref{phiexp}) in (\ref{Vtilde})
\begin{equation}
\tilde{V}\Big|_{\phi=\braket{\phi}}=-\frac{1}{16}F^{2}m_{\phi}^{2}-cF^{2}\chi^{2}+\frac{\lambda}{4!}\chi^{4} \label{VatF}
\end{equation}
Now we can identify the Higgs portal term with the dilatonic Higgs mass as
\begin{equation}
\frac{1}{2}m_{\chi}^{2}=cF^{2}
\end{equation}
In order to produce a light Higgs ($\sim 125\; GeV$) with $\braket{\phi}\sim 1.44\times10^{15}\; GeV$, the coupling constant must be small; that is to say, $c=3.7676\times 10^{-27}.$ \\
Note that $V(\phi)$ is a Coleman-Weinberg type potential \cite{ColemanWeinberg}. Another possible $V(\phi)$ is a $\phi^{4}$ potential, and it would be associated with chaotic inflation. There were previous attempts in the literature where the authors introduced an $SU(5)$ singlet that couples to $H_{5}$ and $\Phi$ to produce inflation without coupling it to gravity \cite{Shafi}, or they proposed that the Higgs is the inflaton \cite{Shapo} \cite{Bezro}. The field $\phi$ can be coupled to gravity as
\begin{equation}
S=\int d^{4}x \sqrt{-g}\Bigg(\frac{1+\kappa^{2}\phi^{2}\xi}{2\kappa^{2}}R-\frac{1}{2}\phi_{;\mu}\phi^{;\mu}-V(\phi) \Bigg)
\end{equation}
where $\kappa^{-1}=M_{p}=2.4 \times 10^{18}\, GeV$ and $\xi$ is the non-minimal coupling. After performing conformal transformation, the potential can be related to the slow-roll parameters in inflation \cite{Kaiser}. $ V(\phi)$ in (\ref{Vphi}) can produce a 60 e-fold inflation, and is consistent with Planck results \cite{Planck} \cite{cosmo}. Therefore, the dilatonic field $\phi$ can be considered as the inflaton that emerges out of conformal invariance at the tree level of leptoquarks in $SU(5)$, and gives a light mass to a dilatonic Higgs through the Higgs portal. 
\section{Conclusion}
In this paper, a dilatonic field $\phi$ emerges as a GB of conformal invariance of leptoquarks in $SU(5)$. The Higgs is considered to be a dilaton $\chi$ as in \cite{Goldberger}. The dilatonic Higgs acquires its mass through the Higgs portal. Hence, this model can produce a light dilatonic Higgs and inflation consistent with Planck results by introducing a dilatonic field that captures conformal symmetry at the tree level of leptoquarks in $SU(5)$.

\paragraph{\textbf{Acknowledgments}}
I would like to thank professors Gordon Semenoff and Nassif Ghoussoub for the useful discussions and support. This research was supported and funded by Ghoussoub's NSERC grant and University of British Columbia.


\begin{thebibliography}{9}
\bibitem{Goldberger} W. D. Goldberger, B. Grinstein, and W. Skiba, Phys.Rev.Lett. 100, 111802 (2008), arXiv:0708.1463 [hep-ph].
\bibitem{Literature}  A. Kobakhidze and S. Liang, arXiv:1701.04927 [hep-ph]; A. Kobakhidze and S. Liang, arXiv:1707.05942 [hep-ph]; S. Arunasalam, A. Kobakhidze, C. Lagger, S. Liang, A. Zhou, arXiv:1709.10322 [hep-ph]
\bibitem{Higgsportal1} T. Binoth and J. J. van der Bij, Z. Phys. C 75, 17 (1997), R. Schabinger and J. D.Wells, Phys. Rev. D 72 (2005) 093007, B. Patt and F. Wilczek, arXiv:hep-ph/0605188.
\bibitem{Higgsportal2} R. Hemping, Phys. Lett. B 379 (1996) 153, and W. -F. Chang, J. N. Ng and J. M. S. Wu, Phys. Rev. D 75 (2007) 115016.
\bibitem{Coleman} S. Coleman, Aspects of Symmetry, Cambridge University Press, 1985.
\bibitem{atheorem1} Z. Komargodski, A. Schwimmer, On renormalization group flows in four dimensions. J. High Energy Phys. 1112, 099 (2011). arXiv:1107.3987 [hep-th].
\bibitem{atheorem2} H. Elvang and T. M. Olson, RG flows in d dimensions, the dilaton effective action, and the a-theorem, JHEP 03 (2013) 034, [arXiv:1209.3424].
\bibitem{ColemanWeinberg}S. R. Coleman and E. J. Weinberg, Phys.Rev. D7, 1888 (1973).
\bibitem{Shafi} Q. Shafi and A. Vilenkin, Phys. Rev. Lett. 52, 691 (1984).
\bibitem{Shapo}F.~L.~Bezrukov and M.~Shaposhnikov, ``The Standard Model Higgs boson as the inflaton,'' Phys.\ Lett.\ B {\bf 659}, 703 (2008) doi:10.1016/j.physletb.2007.11.072 [arXiv:0710.3755 [hep-th]]
\bibitem{Bezro}
F.~Bezrukov, ``The Higgs field as an inflaton,'' Class.\ Quant.\ Grav.\  {\bf 30}, 214001 (2013) doi:10.1088/0264-9381/30/21/214001 [arXiv:1307.0708 [hep-ph]].
\bibitem{Kaiser} D.I. Kaiser, Phys. Rev. D52 (1995) 4295.
\bibitem{Planck}Planck Collaboration, P. A. R. Ade et al., Planck 2015 results. XX. Constraints on inflation, arXiv:1502.02114.
\bibitem{cosmo} G. Panotopoulos, Nonminimal GUT inflation after Planck results, Phys. Rev. D89 (2014), no. 4 047301, [arXiv:1403.0931]; K. Kannike, A. Racioppi and M. Raidal, JHEP 1601 (2016) 035 [arXiv:1509.05423 [hep-ph]]; L. Marzola and A. Racioppi, “Minimal but non-minimal inflation and electroweak symmetry breaking,” JCAP 1610 no. 10, (2016) 010, arXiv:1606.06887 [hep-ph].
\end{thebibliography}
\end{document}